\documentclass[a4paper,11pt]{article}
\usepackage{epsfig}
\usepackage{axodraw}
\newcommand{\be}{\begin{equation}}
\newcommand{\en}{\end{equation}}
\newcommand{\bea}{\begin{eqnarray}}
\newcommand{\ena}{\end{eqnarray}}
\newcommand{\lbl}[1]{\label{eq:#1}}

\newcommand{\lblfig}[1]{\label{fig:#1}}

\newcommand{\rf}[1]{(\ref{eq:#1})}

\newcommand{\fig}[1]{\ref{fig:#1}}

\newcommand{\gapprox}{%
\mathrel{%
\setbox0=\hbox{$>$}\raise0.6ex\copy0\kern-\wd0\lower0.65ex\hbox{$\sim$}}}
\newcommand{\lapprox}{%
\mathrel{%
\setbox0=\hbox{$<$}\raise0.6ex\copy0\kern-\wd0\lower0.65ex\hbox{$\sim$}}}

\newcommand{\Kbar}{\overline{K}}
\newcommand{\mpd}{{m^2_+}}
\newcommand{\mmd}{{m^2_-}}
\newcommand{\mk} {m_K}
\newcommand{\mpi}{m_\pi}

\newcommand{\mkd} {m^2_K}
\newcommand{\mpid} {m^2_\pi}
\def\xlss{\lambda_{s'}}

\def\xls{\lambda_{s}}

\def\Sig{\Sigma}
\def\Del{\Delta}
\def\im{ {\rm Im}\,}
\def\re{ {\rm Re}\,}
\begin{document}
\begin{titlepage}
\begin{flushright}
LPT-ORSAY/06-43\\
\today
\end{flushright}
\begin{center}
{\Large\bf The $K^*_0(800)$ scalar resonance from Roy-Steiner 
representations of $\pi K$ scattering}

\vspace{1cm}

{\large S. Descotes-Genon$^a$ and B. Moussallam$^b$}

\vspace{1cm}

{\sl$^a$\ Laboratoire de Physique Th\'eorique,\\
CNRS/Univ. Paris-Sud 11 (UMR 8627), 91405 Orsay Cedex, France}

{\sl$^b$\ Institut de Physique Nucl\'eaire,\\
CNRS/Univ. Paris-Sud 11 (UMR 8608), 91406 Orsay Cedex, France}

\end{center}
\vfill

\begin{abstract}
We  discuss the existence of the light scalar meson $K^*_0(800)$ 
(also called $\kappa$) in a rigorous way, by showing the presence of
a pole in the $\pi K\to \pi K$ amplitude on the second Riemann sheet.
For this purpose, we study the domain of validity of two classes of Roy-Steiner
representations in the complex energy plane. We prove that one of them is
valid in a region sufficiently broad in the imaginary direction.
From this representation, we compute the $l=0$ partial wave in the complex
plane with neither additional approximation nor model dependence, 
relying only on experimental data. A scalar resonance with strangeness 
$S=1$ is found with the following mass and width: $M_\kappa=658\pm 13$ MeV and 
$\Gamma_\kappa=557\pm 24 $ MeV. 
\end{abstract}
\vfill

\end{titlepage}

One striking aspect of hadron spectroscopy is the extreme scarcity of 
exotics, i.e., states which fail to be understood as either 
$Q\overline{Q}$ or $QQQ$ in the naive quark model. 
It is only recently that several such mesons have been unambiguously
identified in the heavy quark sector (e.g.~\cite{D2317,X3872}). 
While these are very narrow states, one expects from large-$N_c$
considerations~\cite{thooft} that many exotic mesons, on the contrary,  
should be rather wide, which makes them  difficult to be 
singled out experimentally.
From the theoretical point of view, resonances can be defined in a robust and
process-independent way without assumptions on the value of the width, as 
a pole in the $S$ matrix on the second Riemann sheet with
respect to the elastic cut (e.g.~\cite{taylor}). 
In order to locate wide resonances in a reliable
way, one must determine the value of $S$ matrix elements in the complex energy
plane, which requires a careful exploitation of analyticity
properties in association with available experimental data.

In the light quark sector, the scalar mesons lighter than 1 GeV have
been suspected to be exotics for a long time~\cite{jaffe}. 
In this context, it is important to confirm the
existence of the lightest ones, 
namely the $f_0(600)$  with strangeness $S=0$ and 
the $K^*_0(800)$ with $S=1$, also familiarly called $\sigma$
and $\kappa$. 
New indications on the presence of these resonances have been reported
based on the data of
the E791~\cite{E791a,E791b} and BES collaborations~\cite{BES}
(see also~\cite{bugg05,bugg06}) concerning the decays $D^+\to\pi^-\pi^+\pi^+$,
$D^+\to K^-\pi^+\pi^+$ and $J/\Psi\to \pi^+\pi^-\omega$,  
$J/\Psi\to K^+\pi^- K^+\pi^-$ respectively.
Conclusions were drawn from fits to the corresponding Dalitz plots with
Breit-Wigner-like parametrisations. 
In such parametrisations, however, the presence of a pole is assumed
from the start and the description of 
the amplitude in the complex plane is afflicted by
well-known blemishes (spurious poles, absence of left-hand cuts\ldots).
In this paper, we address the question of the existence
of a pole corresponding to the $K^*_0(800)$ without relying on such
approximations. Should this pole exist, one ought to be able to locate it 
in the amplitudes for $D$ or $J/\Psi$ decays
as well as in the amplitude for elastic $\pi K$ scattering. 
Recently, the existence of the $\sigma$ meson has been confirmed
in the $\pi\pi$ scattering amplitude and its mass and width
have been determined quite accurately~\cite{caprini} and we are
following the same kind of method.

The elastic $\pi\pi$- and $\pi K$-scattering amplitudes enjoy rather unique
properties because pions and kaons are the lightest particles
in the QCD spectrum. The analytic structure of the amplitudes is
simple, free from anomalous thresholds, 
and elastic unitarity holds in both direct and crossed channels
in the low-energy region. An additional useful 
property of the $S$ matrix element for elastic scattering is that
a resonance  manifests itself not only as a {\it pole} 
on the second Riemann sheet, but also as a {\it zero} on the first sheet. 

Earlier works performing extrapolations of the $\pi K$ scattering
amplitude in the complex plane have often relied on approximations and
sometimes involved model-dependent hypotheses.
Cherry and Pennington~\cite{penncherr} have used a method based on conformal
mapping and stabilised fit to the data~\cite{pisut}. Ignoring the contributions
from the left-hand cuts, they were inconclusive about the presence of a pole
with $\re(M)< 0.83$ GeV but ruled out a pole at higher mass. Their result
apparently contradicts the earlier claim by Ishida et al.~\cite{ishida} who,
using a naive Breit-Wigner parametrisation of the $\pi K$-scattering data,
found a pole with $\re(M)\simeq 0.88 $ GeV.   In refs.~\cite{zheng1,zheng2}
a novel dispersive representation of the partial wave $S$ matrix was developed,
which also satisfies the elastic unitarity relation at low energy by 
construction. 
An approximation made by these authors consists in using chiral 
perturbation theory
(ChPT) at order $p^4$ in order to compute the discontinuities along the
left-hand cuts (we will comment on the validity of this approximation 
in sec.~1.3). This dispersive construction does yield a $\kappa$ resonance
pole in the $S$ matrix. 
Models for the scattering amplitudes of pseudoscalar mesons have been proposed
by starting from their chiral expansions at leading or next-to-leading order,
later improved by applying a unitarisation ansatz (see e.g.~\cite{ollerosetrev}
for a recent review on this subject). The $K^*_0(800)$ resonance has been
discussed within this framework in refs.~\cite{jop00,pelaez04}. Arguments based
on unitarity bounds applied to a tree-level construction of the amplitude
have been proposed in ref.~\cite{black}.

In this paper, we will show that the existence of
the $K^*_0(800)$, corresponding to a pole in the $S$ matrix, can be established
using \emph{a)} the available experimental data and \emph{b)} general 
properties of analyticity, unitarity and crossing symmetry 
of two-body scattering amplitudes. 
We will rely on  dispersive representations of the Roy-Steiner (RS)
type~\cite{RoySteiner} for the
$I={1\over2}$ $\pi K$ $S$-wave amplitude without performing any 
further approximation. In this approach, the unitarity condition 
on the real axis (below the inelastic threshold)
is not automatically satisfied. It is rather implemented as an equation which
must be solved together with those arising from the dispersive representation 
and from the boundary conditions. This procedure yields the phase shifts in
the energy region below $s_{match}\simeq 1$ GeV$^2$. 
More precisely, in this energy range where elastic unitarity is assumed
to hold to a high precision, a set of six equations is derived,
involving $\pi K\to\pi K$ and $\pi\pi\to K\Kbar$ partial waves with $l=0,1$.
The higher partial waves ($l\ge 2$) and the values for higher energies $s\ge
s_{match}$ must be provided as an input. These equations 
were re-analysed recently in ref.~\cite{pauletal} using
all the available data from high-statistics
experiments~\cite{estabrooks,aston,cohen,etkin}
(earlier work can be found in refs.~\cite{bonnier,johannesson}).

Dispersive representations of scattering amplitudes
have a limited range of validity and it is important to check whether
the putative resonance falls within this domain. We will discuss this point
in some detail below. It will turn out that the particular form of RS 
representation which was considered in ref.~\cite{pauletal} on the real 
energy axis is not valid in a sufficiently large domain in the complex 
energy plane.  A variant will be shown adequate, and we will discuss the
existence, position and features of the $K^*_0(800)$ pole in this context.

\section{Two Roy-Steiner representations of $\pi K$ scattering 
and their domains of validity}

\subsection{Fixed-$t$ representation}

$\pi K$ scattering is described by two different amplitudes $F^+(s,t)$ and
$F^-(s,t)$ which are even and odd respectively under $s$-$u$ exchange
($s$, $t$, $u$ being the standard Mandelstam variables). 
We first start with the representation proposed in ref.~\cite{pauletal}. 
Since the discussion is identical for the two amplitudes, 
we focus on $F^+$ and write a dispersion relation at fixed $t$
with two subtractions (as required from the Froissart bound)
\be\lbl{tfixed}
 F^+(s,t)=c^+(t)+{1\over\pi}\int_\mpd^\infty ds'\,\left[
{1\over s'-s}+{1\over s'-u}-{2s'-2\Sigma-t\over \xlss}
\right] \im_s F^+(s',t)\ 
\en 
with
\be
m_\pm=\mk\pm\mpi,\ \Sigma= \mkd+\mpid,\ \lambda_{s'}=(s'-\mpd)(s'-\mmd)
\en
and $\im_s$ denotes the discontinuity along 
the $s$ cut divided by $2i$.
From the LSZ formula~\cite{lsz},
the representation~\rf{tfixed} can be shown to be valid in a finite 
region of $t$ in a rigorous way~\cite{mandelstam,martin}. 
The range of validity is determined by the possibility to define the
discontinuity function $\im_s F^+(s',t)$ in the whole integration
region of $s'$ through the partial wave expansion
\be\lbl{legserie}
\im_s F^+(s',t)\equiv 16\pi \sum_l (2l+1)\ \im_s f^+_l(s')\ P_l(z(s',t) )
\en
where the argument of the Legendre polynomial is given by
\be
z(s',t)= 1+{2 s' t\over \lambda_{s'} } .
\en
As shown by Lehmann~\cite{lehmanellipse}, the series of Legendre 
polynomials~\rf{legserie} converges when $z(s',t)$ lies inside an ellipse 
whose focal points are located at $z(s',t)=\pm1$ and whose boundary touches
the nearest singularity of $\im_s F^+(s',t)$. 
\vskip 0.5cm
\begin{figure}[h]
\centering
\SetScale{0.25}
\begin{minipage}{6.0cm}
\begin{picture}(500,75)(0,0)
\SetWidth{3.0}
\DashLine(0,0)(500,0){10}
\DashLine(0,300)(500,300){500}
\DashCArc(500,150)(427,159.44,200.55){10}
\DashCArc(800,150)(427,159.44,200.55){10}
\DashCArc(-300,150)(427,-20.56,20.56){10}
\DashCArc(0,150)(427,-20.56,20.56){10}
\CCirc(400,300){20}{Black}{Yellow}
\CCirc(400,0){20}{Black}{Yellow}
\CCirc(100,300){20}{Black}{Yellow}
\CCirc(100,0){20}{Black}{Yellow}
\Text(-20,0)[]{$\pi$}
\Text(-20,75)[]{$K$}
\end{picture}
\end{minipage}\begin{minipage}{5cm}
\SetScale{0.25}
\begin{picture}(500,75)(0,0)
\SetWidth{3.0}
\DashLine(0,0)(100,0){10}
\DashLine(0,300)(100,300){500}
\DashLine(400,0)(500,0){10}
\DashLine(400,300)(500,300){500}
\DashLine(100,0)(100,300){10}
\DashLine(400,0)(400,300){10}
\DashCArc(250,700)(427,-110.56,-69.44){10}
\DashCArc(250,400)(427,-110.56,-69.44){10}
\DashCArc(250,400)(427,-110.56,-69.44){10}
\DashCArc(250,-400)(427,69.44,110.56){10}
\DashCArc(250,-100)(427,69.44,110.56){500}
\CCirc(400,300){20}{Black}{Yellow}
\CCirc(400,0){20}{Black}{Yellow}
\CCirc(100,300){20}{Black}{Yellow}
\CCirc(100,0){20}{Black}{Yellow}
\Text(62.50,90)[]{$q_1$}
\Text(62.50,60)[]{$q_2$}
\Text(62.50,15)[]{$q_3$}
\Text(62.50,-15)[]{$q_4$}
\end{picture}
\end{minipage}
\vskip 0.5cm
\caption{\sl Diagrams showing the contributions of the lightest particles in 
the $s$-channel and the $t$-channel unitarity relations for the $\pi K$
scattering amplitude. Dashed lines correspond to $\pi$ mesons, whereas solid
lines represent $K$ mesons.}
\lblfig{stdiags}
\end{figure}
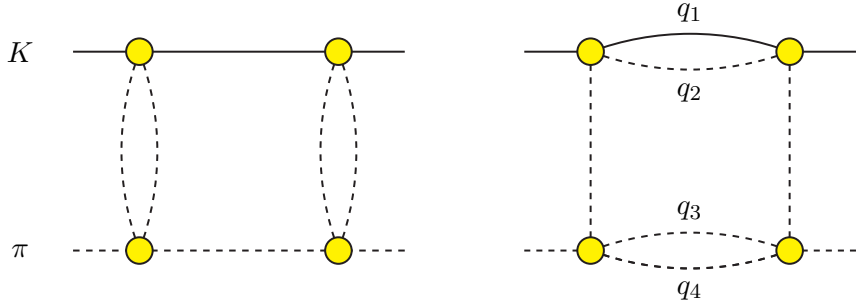

We assume that the scattering amplitude satisfies
Mandelstam's double spectral representation~\cite{mandelstam2}, so that
the nearest singularity is given by the boundary
of the support of the double spectral functions $\rho_{st}$, $\rho_{us}$.
We recall that these boundaries are generated by considering the lightest 
contributions in the unitarity relations. For instance, in the case 
of $\pi K$ scattering, the $st$ boundary comes from the contributions
illustrated in fig.~\fig{stdiags}.
It may be written as $t= T_{st}(s)$
with\footnote{The formulae given in ref.~\cite{pauletal} correspond to
pion-nucleon scattering with $m_N$ simply replaced by $m_K$ which is incorrect.
Using the right expressions yields only small numerical modifications to the
domains of validity on the real axis quoted in ref.~\cite{pauletal}.}
\bea
&& T_{st}(s)= 16 m_\pi^2 +{64 m_\pi^4 s \over (s-(m_K-m_\pi)^2)
(s-(m_K+m_\pi)^2) } \ {{\rm when}\ s \le s_0} 
\nonumber\\
&& T_{st}(s)=  4 m_\pi^2 +{32 m_\pi^3 (m_K+m_\pi)\over (s-(m_K+3m_\pi)^2) } 
\ {{\rm when}\ s \ge s_0 }
\ena
with $s_0= m_K^2 +4 m_K m_\pi + 5 m_\pi^2 +2\mpi 
(5 m_K^2 +12 m_K m_\pi+ 8 m_\pi^2)^{1\over2} $.

The expression for the boundary associated with the $us$ spectral function
can be put in the form $t=T_{us}(s)$ with
\bea
&& T_{us}(s)= {-16\mpid\mpd\over s- (\mk+\mpi)^2} 
-s + m_+ (\mk-7\mpi)
\ {{\rm when}\ s \le s_0} 
\nonumber\\
&& T_{us}(s)= {-16\mpid\mpd\over s- (\mk+3\mpi)^2} 
-s + (\mk-\mpi)^2
\ {{\rm when}\ s \ge s_0}\ . 
\ena
In the complex $t$ plane, the dispersive representation~\rf{tfixed} is
restricted by the $st$ double spectral function to a domain of validity
with the following boundary, expressed in polar coordinates:
\be \lbl{polart}
T(\theta)=\min_{\mpd\le s'\le\infty} T_{s'}(\theta)\quad {\rm with} \quad
T_{s'}(\theta)= {T_{st}(s') (\xlss +s' T_{st}(s'))\over\xlss \cos^2
{\theta\over2} + s' T_{st}(s') }\ .
\en
A RS representation is generated by projecting eq.~\rf{tfixed} on
the $l=0$ partial wave,
\be\lbl{project}
f^+_0(s)= {s\over16\pi\lambda_s}\int_{-{\lambda_s\over s} }^0 dt\, F^+(s,t)\ .
\en
This projection can be performed only if the segment of integration remains
inside the region of validity~\rf{polart}. The boundary of the
domain of validity of the RS representation in the $s$-plane 
is therefore obtained, in parametric form, by solving 
\be
\lambda_s + s\ T(\theta) \exp(i\theta) =0\ .
\en
\begin{figure}[h]
\centering
\includegraphics[width=11cm]{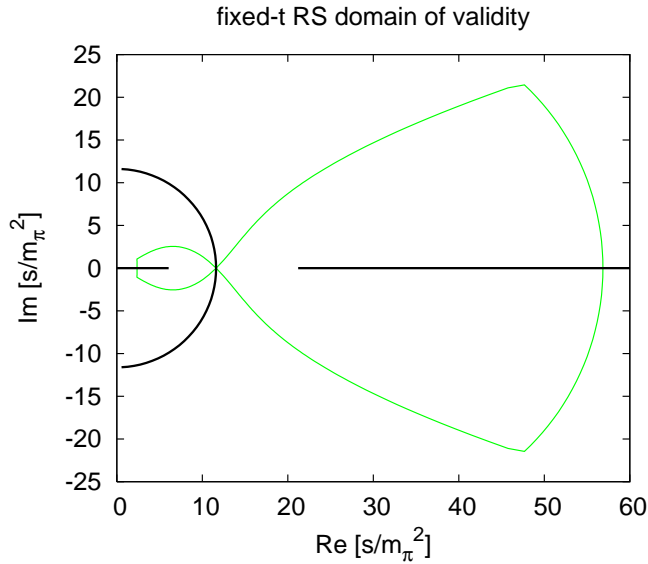}
\caption{\sl  Domain of validity of the Roy-Steiner representation 
based on fixed-$t$ dispersion relation. The energy variable $s$ is 
expressed in units of $m_{\pi^+}^2$.}
\lblfig{tboundst}
\end{figure}
The result is displayed in fig.~\fig{tboundst} where the two 
cuts along the real axis as well as the circular cut of the partial wave
amplitude are also drawn. As can be seen on this figure,
the validity region of the Roy-Steiner representation based on fixed-$t$
dispersion relation gets squeezed when $\re(s)$ is close to the
$\pi K$ threshold, which makes it a priori unfit to search for 
a wide resonance like the $\kappa$. 

\subsection{Fixed-$us$ representation}
Let us now investigate a second kind of dispersion relation, sometimes
called hyperbolic, in which the 
product $us$ is kept fixed~\cite{pauletal}. Setting $us=b$, we get 
a representation for $F^+(s,t)$ of the following form:
\bea\lbl{bfixed}
&& F^+(s,t_b(s))= f^+(b) + t_b(s) h^+(b) 
\nonumber\\
&&   \phantom{ F^+(s,t_b(s))}
+ 
{1\over\pi} \int_\mpd^\infty ds'\left[  
 {2s'-2\Sig+t_b(s)\over (s'-s)(s'-b/s)}
-{2s'-2\Sig-t_b(s)\over s^{'2}-2\Sig s'+b} \right]\im_s F^+(s',t_b(s'))
\nonumber\\
&&   \phantom{ F^+(s,t_b(s))}
+ {t_b(s)^2\over \pi}\int_{4\mpid}^\infty 
{dt'\over t^{'2}(t'-t_b(s))}\im_t F^+(s'_b(t'),t')
\ena
with
\be
t_b(s)= 2\Sig-s -{b\over s},\quad
s'_b(t')= {1\over2} \left(2\Sig-t' +\sqrt{ (2\Sig-t')^2-4b}\right)\ .
\en
Expanding the discontinuities in partial waves and projecting 
the whole representation
on the $l=0$ partial wave yields a RS 
representation which we denote ${\rm RS_b}$ and which is different from the
fixed-$t$ representation considered earlier. 

Let us now consider the domain of validity of this new representation. We must
ensure that the discontinuity functions $\im_s F^+(s',t_b(s'))$ and
$\im_t F^+(s'_b(t'),t')$ are defined inside the $s'$ and the $t'$ integration 
ranges, once these functions are expanded on $\pi K\to\pi K$
and $\pi\pi\to K\Kbar$ partial waves respectively. As done before,
we consider each Mandelstam boundary ($st$ and $us$) 
and we determine the region for the parameter $b$
inside which the representation~\rf{bfixed} is valid. Let us denote
$B(\theta)$ the description (in polar coordinates) for the boundary of
such a region.
The $S$-wave  component of this representation is then taken through
\be
f^+_0(s)= {1\over16\pi\lambda_s}\int_{\Delta^2}^{(2\Sigma-s)s} db\, F^+\left(s,2\Sig-s -{b\over s}\right)\ .
\en
The segment of integration (i.e., its end at $(2\Sigma-s)s$) must 
remain within the region of validity in $b$,
so that the boundary in the $s$ plane for the ${\rm RS_b}$ representation is
obtained as a solution to
\be
s^2-2\Sig s + B(\theta)\exp(i\theta)=0\ .
\en

\begin{figure}[ht]
\centering
\includegraphics[width=12cm]{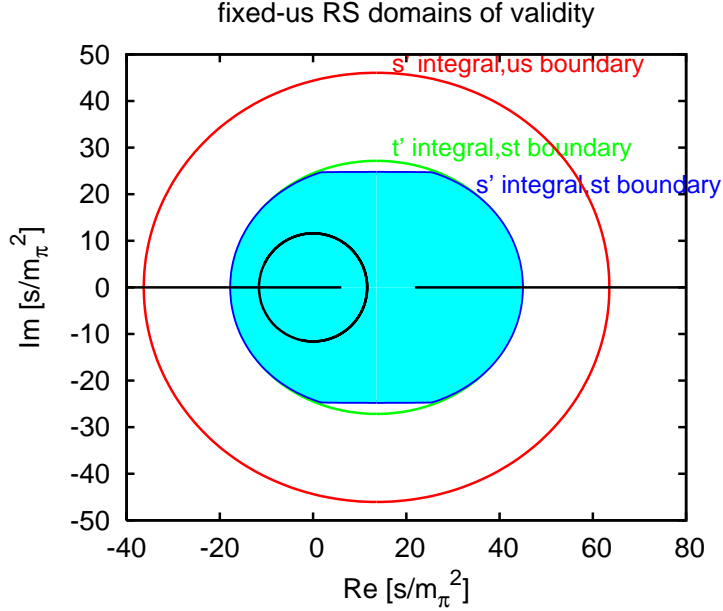}
\caption{\sl Domains of validity associated with the $s'$ and $t'$
integrals in the fixed-$us$ representation \rf{bfixed} and resulting from the
conditions that the Lehmann ellipses do not touch the $st$ or the $us$
Mandelstam boundaries. }
\lblfig{usvalid}
\end{figure}

The domains of validity which result from the consideration of the
$s'$ and $t'$ integrals are shown in fig.~\fig{usvalid}. In the case
of the $t'$ integral, the $st$ Mandelstam boundary is the only one relevant.
In the case of the $s'$ integral, one must consider both the $us$ and
the $st$ Mandelstam boundaries. The last domain is included into all
the others and therefore defines the region in the complex plane 
where the ${\rm RS_b}$ representation is valid. 
The shape of this domain is quite different from fig.~\fig{tboundst}
corresponding to the fixed-$t$ RS representation.
The latter exhibits a more extended validity along the real axis, whereas the
former is significantly broader along the imaginary direction. Indeed,
the domain of validity for ${\rm RS_b}$ extends up to $\im(s)\simeq 0.39$
GeV$^2$  when $\re(s)$ is close to the threshold, which will turn out to be
sufficient for the $K^*_0(800)$ resonance.

\subsection{The ${\rm RS_b}$ representation of the scalar partial wave}

Let us give more details on the representation of the $f_0^{1\over2}$ 
partial wave.
The functions $f^+(b)$ and $h^+(b)$ which appear in eq.~\rf{bfixed}
have been determined in ref.~\cite{pauletal}. 
Carrying out the projection of the amplitudes $F^+$ and $F^-$ in the 
form~\rf{bfixed}, we obtain the
$\pi K\to\pi K$ amplitude of isospin $I=1/2$ for the partial wave $l=0$, 
\bea\lbl{rsbrep}
&& f_0^{1\over2}(s) =
{1\over2} m_+ a_0^{1\over2} + {1\over12} m_+ (a_0^{1\over2}
-a_0^{3\over2}) { (s-\mpd)(5s +3\mmd)\over (\mpd-\mmd)\, s }
\nonumber\\
&&\phantom{f_0^{1\over2}(s)}
+{1\over\pi}\int_\mpd^\infty ds' \sum_{l=0}^\infty \left\{
K_{0l}^{1\over2} (s,s') \im f^{1\over2}_l(s') +
K_{0l}^{3\over2} (s,s') \im f^{3\over2}_l(s') 
\right\}
\nonumber\\
&&\phantom{f_0^{1\over2}(s)}
+{1\over\pi}\int_{4\mpid}^\infty dt' \sum_{l=0}^\infty \left\{
   K_{0 2l}^0(s,t')\im g^0_{2l}(t') 
+  K_{0 2l+1}^1(s,t')\im g^1_{2l+1}(t') 
\right\}\ ,
\ena
where $a_0^I$ denote the scattering lengths. 
This is the key expression which will enable us to compute the
amplitude $f_0^{1\over2}(s)$ for complex values of $s$.
The first few of the kernels which act on the $\pi K\to\pi K$
partial waves $f_l^I(s')$ read
\bea
&& K_{00}^{1\over2} (s,s') ={1\over s'-s} -{1\over3} L(s,s')
-{ 4s(s'+2s-3\Sig)-3\xls\over 6 s\xlss }
\nonumber\\
&&  K_{01}^{1\over2} (s,s') =
  \left(1+ {2(s s'-\Delta^2)\over\xlss}\right) L(s,s')
-{ 2s(s'+s) +3\xls\over 2 s \xlss}
\nonumber\\
&&  K_{00}^{3\over2} (s,s') ={4\over3} L(s,s') 
-{8s(s'-s) + 3\xls \over 6 s\xlss }
\ena
with
\be
L(s,s')= {s\over\xls}\left(\log( s'+s -2\Sig)-
\log \left(s'- {\Delta^2\over s}\right)\right)
\en
and
\be
\Delta=\mkd-\mpid, \quad \Sigma= \mkd+\mpid\ .
\en
We quote a few of the kernels which act on the $\pi\pi\to K\Kbar$
partial waves $g_l^I(t')$
\bea
&& K_{00}^0(s,t')={1\over\sqrt3}\left\{ \widehat L(s,t')-{1\over t'}\right\}
\nonumber\\
&& K_{01}^1(s,t')={3\sqrt2\over4}\left\{ (2s-2\Sig+t')\widehat L(s,t')
+{-5s^2+2\Sig s +3\Del^2\over 4t' s}-1 \right\}
\ena
with
\be
\widehat L(s,t')={s\over\xls}\log\left(1+ {\xls\over s t'} \right)\ .
\en
A few comments are in order at this point:
\begin{itemize}
\item[--] In the formula~\rf{rsbrep}, the integrands are evaluated using
the description of $\pi K$ scattering (and its crossed channel) 
along the real axis obtained by solving Roy-Steiner 
equations~\cite{pauletal}.

More precisely, whenever the integration variables $s'$, $t'$ are larger
than approximately 1 GeV$^2$, the imaginary parts $\im f_l^I(s')$, 
$\im g_l^I(t')$ are taken from fits to the experimental data 
(see~\cite{pauletal} for more details).
In practice, experimental information 
is available for values of $l$ up to $l=5$ and in a range of
energies up to $s'_{max}\simeq t'_{max}\simeq 6 $ GeV$^2$. The 
integrals involved here converge quickly and we restrict ourselves 
to values of $s$ such that $\vert s\vert \lapprox 1$ GeV$^2$. 
We can conclude that we only need qualitative 
estimates for the imaginary parts in the higher integration region.
For this purpose, the simple Regge pole models used in ref.~\cite{pauletal} 
are appropriate.

In the lower parts of the integration ranges, $\im f_l^I(s')$, 
$\im g_l^I(t')$ with $l=0,1$ are taken from the solutions of the RS equations
computed in~\cite{pauletal}. The scattering lengths $a_0^{1\over2}$
and $a_0^{3\over2}$ were also obtained from these solutions
\be
a_0^{1\over2}= (0.224\pm0.022) m_\pi^{-1},\quad
a_0^{3\over2}= (-0.448\pm0.077)10^{-1} m_\pi^{-1}\ .
\en
When $s$ is on the real axis
with $\mpd \le s\le s_{match}$, we have  verified that $f_l^I(s)$
with $l=0,1$ as given from the ${\rm RS_b}$ representation do satisfy the
unitarity relation to a good approximation. In other terms, these amplitudes
satisfy both the RS and the ${\rm RS_b}$ equations.

\item[--] 
The discontinuities of the amplitude $f_0^{1\over2}(s)$ are
generated from the pole $1/(s'-s)$ and from the logarithmic functions 
$L(s,s')$ and $\widehat L(s,t')$ present in the kernels. 
For illustration, let us consider the discontinuity along the circular cut. 
Across a point $s=\Delta\exp(i\theta)$ belonging to the circle, 
the discontinuity is easily computed from eq.~\rf{rsbrep} (noticing that
the circular cut is contained inside the domain of validity of this 
representation) to be
\bea
&& f_0^{1\over2}( (\Delta-\epsilon){\rm e}^{i\theta})
 - f_0^{1\over2}( (\Delta+\epsilon){\rm e}^{i\theta})= 
{ 2i\over 4\mpid +4(\mkd-\mpid)\sin^2{\theta\over2} }\times
\\
&& \quad
\int_{4\mpid}^{4\mpid+4\Delta\sin^2{\theta\over2}} dt' \left\{
{\sqrt3\over3} \im g_0^0(t') -{3\sqrt2\over4}(-2\Delta {\rm e}^{i\theta}
+2\Sig -t') \im g_1^1(t')+\cdots \right\} \ .
\nonumber
\ena

This expression highlights the region of the circle where one is allowed to 
compute the discontinuity using ChPT, i.e.,  replacing the imaginary
part of the $\pi\pi\to K\bar{K}$ partial waves by their chiral expansions
(as done in~\cite{zheng1,zheng2}). The chiral expansion is expected to 
converge in a range $\sqrt{t'}\lapprox 0.5$ GeV: this corresponds 
to the forward portion of the circle with $-52 \lapprox \theta \lapprox 
52$ degrees. Similar expressions can be derived without difficulty for the 
discontinuities along the left-hand cuts on the real axis. 
Again, it is easy to see that ChPT is
applicable over a small portion  of the cut and not, in particular, close
to the point $s=0$.
\end{itemize}
\begin{figure}[ht]
\centering
\includegraphics[width=11cm]{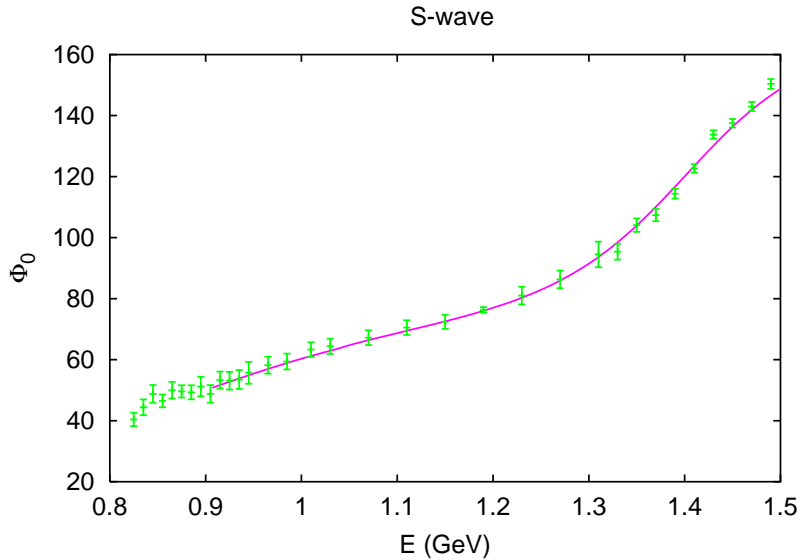}
\caption{\sl Experimental values for the S-wave phases of the amplitude
for charged $\pi K$ scattering measured in ref.~\cite{aston}.}
\lblfig{l=0phase}
\end{figure}

\section{The lightest scalar resonance in $\pi K$ scattering}
Phase-shift analyses for $\pi K\to\pi K$ scattering have been performed 
based on high-statistics production experiments in 
refs.~\cite{estabrooks,aston}. 
For instance, fig. \fig{l=0phase} recalls the $l=0$ phase of
the amplitude $\pi^+ K^-\to \pi^+ K^-$ given in ref.~\cite{aston}. 
The phase displays a typical resonance-like behaviour in connection with
the $K^*_0(1340)$ but no similar behaviour occurs in relation with 
the lighter $K^*_0(800)$. In fact, it is difficult 
to immediately draw any definite conclusion from these data,
since the experimental information does
not quite cover the energy region which would be of interest for the 
$K^*_0(800)$ resonance. 
In order to decide on the existence of this resonance
one must combine the experimental data with theoretical 
constraints. Roy-Steiner representations provide such constraints by
embedding information on the analyticity structure,
unitarity along the real axis as well as crossing symmetry for 
the $\pi K$ scattering amplitude. As discussed
above, one such representation yields $f_0^{1\over2}(s)$ 
in the complex region of $s$ shown in fig.\fig{usvalid}, which lies on the
first Riemann sheet with respect to all the cuts. 
Let us recall here the well-known result that the elastic
$S$ matrix element
\be
S_l^{1\over2}(s)= 1- 2{\sqrt{(\mpd-s)(s-\mmd)}\over s} f_l^{1\over2}(s)
\en
exhibits a resonance as a zero on the first sheet as well
as as a pole on the second sheet. This fortunate property stems from the
unitarity relation  which can be recast, using the analyticity properties, as
an equation between the values of the amplitude on both sides of the cut
\be
f_l^{1\over2}(s-i\epsilon)- f_l^{1\over2}(s+i\epsilon)=
2i {\sqrt{(s-\mpd)(s-\mmd)}\over s} f_l^{1\over2}(s+i\epsilon)
f_l^{1\over2}(s-i\epsilon)\ .
\en  
This relation holds for real values of $s$ along the elastic cut
below the first inelastic threshold. It can be   
translated into a relation for the $S$ matrix
\be\lbl{sunits}
S_l^{1\over2}(s+i\epsilon) S_l^{1\over2}(s-i\epsilon)=1\ .
\en
Introducing a variable $z=-\sqrt{\mpd-s}$ which maps the first sheet of the
$s$ plane onto the upper part of the $z$ plane, we can rewrite eq.~\rf{sunits}
as
\be\lbl{sunitz}
S_l^{1\over2}(z) S_l^{1\over2}(-z)=1\ .
\en
\begin{figure}
\centering
\includegraphics[width=12cm]{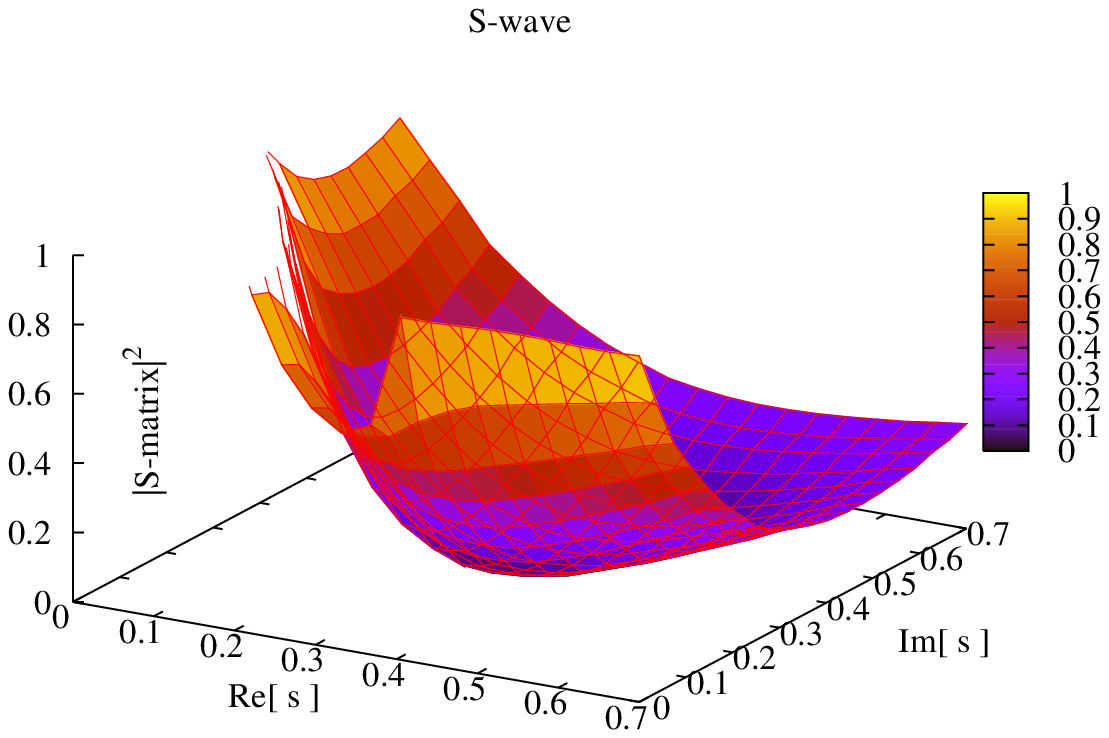}
\caption{\sl Plot of $\vert S_0^{1\over2}(s) \vert^2$
for complex values of $s$ (in units of GeV$^2$), 
computed from the ${\rm RS_b}$ representation~\rf{rsbrep}.}
\lblfig{smat2S}
\end{figure}
\begin{figure}
\centering
\includegraphics[width=12cm]{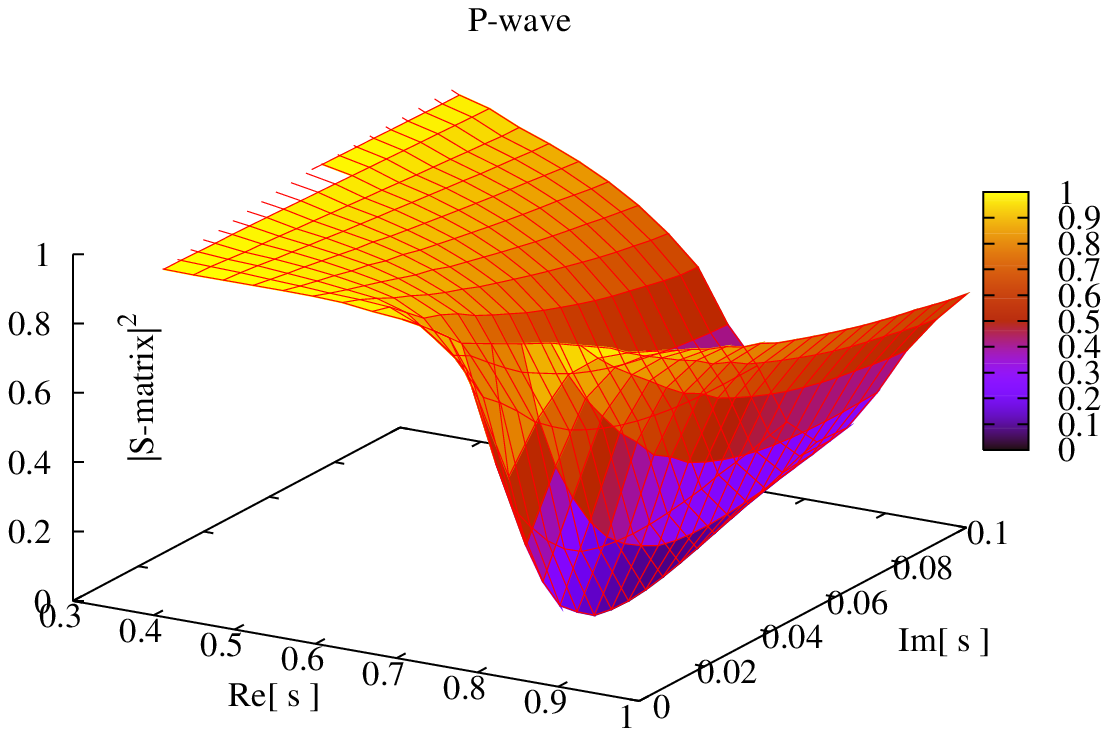}
\caption{\sl  Same as fig.~\fig{smat2S} showing $\vert S_1^{1\over2}\vert^2$.}
\lblfig{smat2P}
\end{figure}
The relation~\rf{sunitz} holds on a finite portion of the positive real
axis. By analytic continuation, it must also hold everywhere in the 
complex $z$ plane. This relation immediately shows that a resonance
pole $z_0$ on the second Riemann sheet $[\im(z_0)<0]$ is automatically 
associated to a zero at $-z_0$, which lies on the first sheet. 
Computing  $S_0^{1\over2}(s)$ from the ${\rm RS_b}$ 
representation described above for the central values
of our experimental input, we find that it does have a  zero, 
$S_0^{1\over2}(s_0)=0$ with
\be\lbl{s0}
s_0 = 0.356 +i\cdot 0.366\ {\rm GeV}^2\ .
\en

The global shape of the $S$ matrix for complex values of $s$
is illustrated in fig.~\fig{smat2S},
which displays the squared modulus of $S_0^{1\over2}(s)$ resulting from
our computation. The figure shows that the
modulus is constant and equal to one over a portion of the real axis 
(in accordance with unitarity) 
and drops when one leaves this axis, eventually becoming zero at $s=s_0$. 
We notice the similarity of the global behaviour of the $S$ matrix 
with the case of an ordinary narrow resonance. 
Indeed, fig.~\fig{smat2P} shows 
the squared modulus of the $P$-wave $S$ matrix 
computed using the same apparatus,
which exhibits the well-known $K^*(890)$ resonance as a zero. 
According to these results, the existence of the $K^*_0(800)$ scalar 
resonance is established on the same footing as that of the 
vector $K^*(890)$ resonance.

However, one may highlight a difference between the two situations that 
is illustrated in fig.~\fig{zerolines}. In the complex $s$-plane
are drawn the two lines $L_R$ and $L_I$ defined as the curves along
which the real and imaginary parts of the $l=0$ $S$ matrix vanish respectively
(the point $s_0$ corresponds to the intersection of these two lines). The 
line $L_R$ starts from the real axis at the point where the phase shift
is equal to $\pi/4$ (since $\re(S_0^I)=\cos(2\delta_0^I)$). 
In the case of an ordinary resonance, the line
$L_I$ would start from the real axis at the point where the phase shift
reaches $\pi/2$, whereas it starts from a point situated slightly below
the elastic cut in the case of the $K^*_0(800)$.
\begin{figure}   
\centering
\includegraphics[width=11cm]{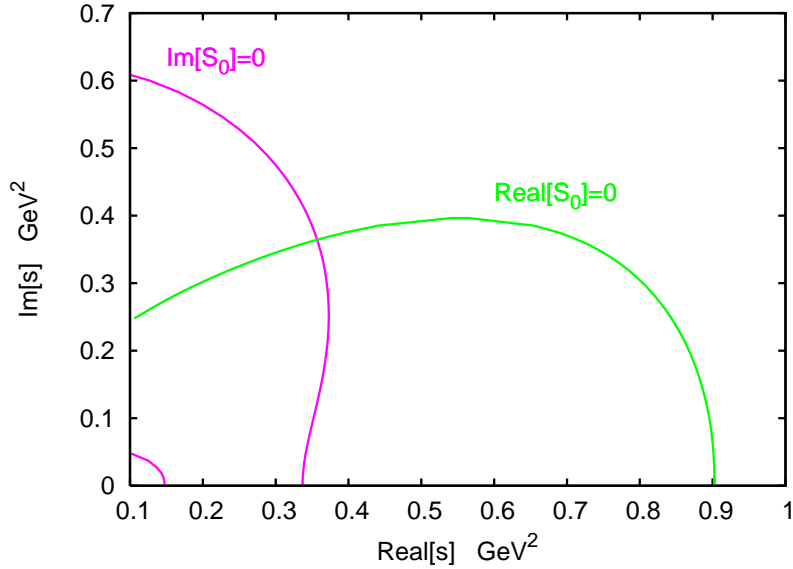}
\caption{\sl Lines $L_R$ and $L_I$ in the complex $s$-plane 
along which $\re{ S_0^{1\over2}}(s)$ and
$\im{ S_0^{1\over2}}(s)$ vanish respectively.}
\lblfig{zerolines}
\end{figure}

As stated earlier, the point $s_0$ is located inside the domain of validity
of the ${\rm RS_b}$ representation. This is illustrated in fig.~\fig{sellipse}
which shows the one-sigma error ellipse on $s_0$ 
computed by varying the parameters describing our input data (see
ref.~\cite{pauletal} for more detail). In addition the figure shows that $s_0$
is located at about the same distance from the physical cut as from the
circular cut. This feature confirms that a representation
of the amplitude accounting for the left-hand cuts correctly is needed in order
to determine $s_0$ in a reliable way. The mass and width of the 
$\kappa$ resonance, as defined from the square root of $s_0$, 
$M_\kappa+i\cdot\Gamma_\kappa/2 = \sqrt{s_0}$, are then found to be
\be
M_\kappa= 658 \pm 13 \ {\rm  MeV}\,,\quad
\Gamma_\kappa=557 \pm 24 \ {\rm  MeV}\ .
\en
\begin{figure}
\centering
\includegraphics[width=12cm]{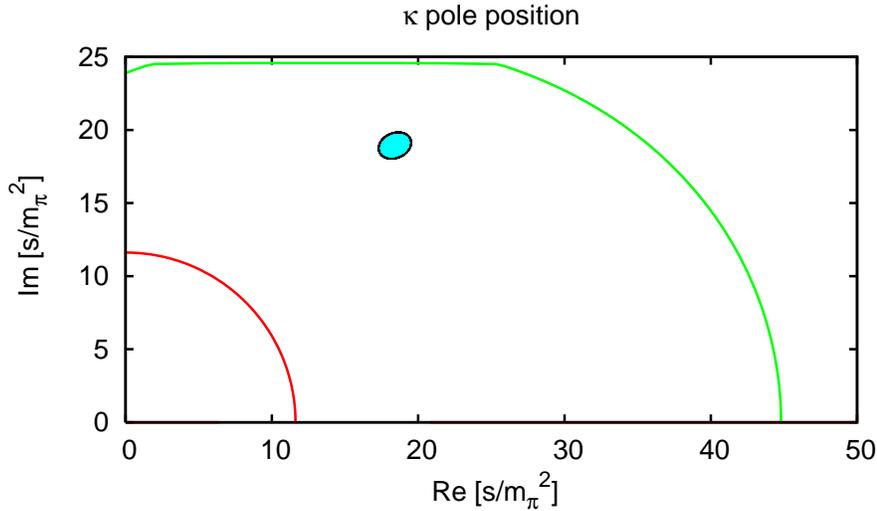}
\caption{\sl Position of the $\kappa$ pole $s_0$ and its one-sigma error
ellipse (in units of $m_\pi^2$). 
We also show the boundary of the region of validity of the 
${\rm RS_b}$ representation and the left-hand cuts of the amplitude.}
\lblfig{sellipse}
\end{figure}
The errors are rather small and of the same size as the errors
affecting the $\sigma$-meson mass and width as obtained in 
ref.~\cite{caprini}. This reflects the good quality of the experimental
data used as input (see e.g. fig.~\fig{l=0phase}) which is exploited in an
optimal way. However, since the
point $s_0$ is not located very from the boundary of the region of 
validity, one may wonder whether a significant uncertainty might be
introduced by truncating the partial-wave series in the various 
integrands. As a matter of fact, this is unlikely, because 
the $RS$ representations are expected not to break down
in an abrupt way when the validity boundary is crossed. Let us consider
the diagrams associated with the Mandelstam regions 
for $\pi K$ scattering used to determine the validity boundary. 
The $st$ boundary corresponds to setting the invariants
$(q_1+q_2)^2$ and $(q_3+q_4)^2$ to their threshold values, see 
fig.~\fig{stdiags}. The actual dispersive representation of the amplitude 
is expected to involve weight functions that should
be suppressed at threshold (in particular due to chiral symmetry) and that 
should be peaked for values of $(q_1+q_2)^2$ and $(q_3+q_4)^2$ in the resonance
region (corresponding to an effectively more distant Mandelstam
boundary). Because of these effects, crossing the validity boundary should
affect the accuracy of the Roy-Steiner representations in a mild way : one
has to venture much deeper into the complex plane to notice a significant
breakdown of the dispersion relations.   
As an illustration, we have computed the position of 
the zero, $s_0$, using the fixed-$t$ Roy-Steiner representation. Using
this dispersion relation significantly outside of its strict domain of 
validity, we have found a difference of only 0.5~\% in the pole position
in comparison with the result from the $\rm{RS_b}$ representation.

In table 1 we summarise the results of a few other determinations
of the $K^*_0(800)$ resonance parameters in the recent literature.
These are derived from input experimental data on $\pi K$ scattering, except
for the result of Aitala et al.~\cite{E791b} which is based on $D\to K\pi\pi$
decays and the one from Bugg~\cite{bugg06} who uses the same data combined
with BESS II data on $J/\psi\to K^*(890)K\pi$. 
Our results are compatible with those of~\cite{zheng1,zheng2} 
who have also employed dispersive methods. The mass which we find is lighter 
than in previous calculations. A similar effect was observed in 
ref.~\cite{caprini} in the case of the $\sigma$ and it was traced to a more
complete  treatment of the left-hand cuts in Roy-type representations. 
\begin{table}[h]
\centering
\begin{tabular}{|l|l|l|}\hline\hline
\ & $M_\kappa$ (MeV)& $\Gamma_\kappa$ (MeV)\\ \hline
This work                   & $658\pm 13 $      & $557 \pm 24$   \\ \hline
Zhou, Zheng~\cite{zheng2}   & $694\pm 53 $      & $606\pm89$     \\
Jamin et al.~\cite{jop00}   & $708$             & $610$         \\ 
Aitala et al.~\cite{E791b}  & $721\pm 19\pm43$  & $584\pm43\pm87$\\
Pelaez~\cite{pelaez04}      & $750\pm 18 $      & $452\pm22$     \\
Bugg~\cite{bugg05}          & $750^{+30}_{-55}$ & $684\pm120$   \\
Ablikim et al.~\cite{officialbess}& $841\pm23^{+64}_{-55}$& 
$618\pm52^{+55}_{-87}$ \\
Ishida et al.~\cite{ishida} & $877^{+65}_{-30}$ & $668^{+235}_{-110}$\\
\hline\hline
\end{tabular}
\caption{\sl The mass and width of the $K^*_0(800)$ from our work 
and some other recent determinations. Refs.~\cite{E791b,officialbess,
ishida} quote Breit-Wigner parameters from which we 
have computed the corresponding pole positions.}
\end{table}

\section{Summary and outlook}

It is quite likely that many exotic mesons (or baryons) exist in QCD which
are not seen simply because they have a very large width. In the case of
the $\kappa$ meson, we have demonstrated that it is perfectly possible
to prove the existence of such particles by combining experimental data
with some general theoretical constraints. Previously, the same
conclusion was derived in the case of the $\sigma$ 
meson~\cite{caprini}. A major advantage of the methods used here 
and in ref.~\cite{caprini} lies in the control of their
range of validity as one moves away from the physical energy region
into the complex plane. No such control exists for naive parametrisations
of the Breit-Wigner type or even for more sophisticated ones like
chiral-unitarised approaches.

The $\pi K$-scattering matrix in the $S$ wave 
has been computed in the complex energy plane using a Roy-Steiner 
dispersive representation. It is worth noting that
in such a representation, one must inject 
much more experimental information than just the $S$-wave phase shifts  
(such as data on other $\pi K$ and crossed-channel partial waves and the high 
energy behaviour). Moreover, the available $S$-wave 
data does not cover the lower energy range. 
In this region, unitarity provides extra 
information which can be combined with the RS representation to
compensate for the lack of experimental data. 
The combination of experimental and theoretical information
leads to a zero of the $S$ matrix on the first sheet, and therefore a pole on
the second one, which confirms the existence of the $K^*_0(800)$ resonance. 
We have observed that the behaviour of the $S$ matrix when the energy variable
$s$ becomes complex is qualitatively the same as in the case of a narrow
resonance: the $S$ matrix makes no difference between an ordinary and
an exotic meson. 

In ref.~\cite{caprini} the following results were found for the   
$\sigma$ meson
\be
M_\sigma=441^{+16}_{-8} \ {\rm MeV},\quad \Gamma_\sigma=544^{+18}_{-25}\ {\rm
  MeV}.
\en
Comparing $M_\kappa$ and $M_\sigma$ 
suggests that the $\kappa$ meson is the $S=1$ partner
of the $\sigma$ meson. This tends to disfavour the scenario proposed 
by Minkowski and Ochs~\cite{minkochs} in which the $\sigma$ 
contains a sizable glueball component. 
If one formed a nonet by associating together the 
$\sigma$, the $\kappa$, the iso-singlet $f_0(980)$ and 
the iso-triplet $a_0(980)$, 
its mass pattern would be clearly at variance with the usual 
$Q\bar Q$ picture (which is also what is expected in the large $N_c$ limit
of QCD).  In contrast, it would be conspicuously similar to
the pattern predicted by Jaffe from a $Q^2\bar Q^2$ 
picture a long time ago~\cite{jaffe}. The correct values 
for the widths seem more difficult to reproduce in simple quark
models~\cite{Maiani04}.
Many different models, multiplet assignments and interpretations of the light scalar mesons 
have been proposed in the literature (see~\cite{atornqvist} for a review). 
In the future, model-independent information is expected from
lattice simulations of QCD, which start to provide quantitative
predictions on scalar mesons and should give further insights
into this long-standing issue~\cite{lattice}. 

\section*{Acknowledgments}
We would like to thank  D.V. Bugg for encouragement and correspondence.
Instructive conversations with H. Leutwyler and G. Colangelo are also
gratefully acknowledged. 

This work is supported in part by the EU RTN contract
HPPRN-CT-2002-00311 (EURIDICE) and 
the European Community-Research Infrastructure Activity
under the FP6 "Structuring the European Research Area"
programme (HadronPhysics, contract number RII3-CT-2004-506078)

\end{document}